\documentclass[jkps,preprint,fleqn,showpacs,showkeys]{revtex4}
\usepackage[pdftex]{graphicx}
\usepackage{amssymb}
\usepackage{amsmath}
\usepackage{bm}
\usepackage{hyperref}
\begin{document}
\setcounter{page}{0}
\title[]{Start-to-end simulation with rare isotope beam for post accelerator of the RAON accelerator}
\author{Hyunchang \surname{Jin}}
\email{hcjin@ibs.re.kr}
\author{Ji-Ho \surname{Jang}}
\affiliation{Rare Isotope Science Project, Institute for Basic Science, Daejeon 34047, Korea}

\date[]{}

\begin{abstract}
The RAON accelerator of the Rare Isotope Science Project (RISP) has been developed to create and accelerate various kinds of stable heavy ion beams and rare isotope beams for a wide range of the science applications. In the RAON accelerator, the rare isotope beams generated by the Isotope Separation On-Line (ISOL) system will be transported through the post accelerator, namely, from the post Low Energy Beam Transport (LEBT) system and the post Radio Frequency Quadrupole (RFQ) to the superconducting linac (SCL3). The accelerated beams will be put to use in the low energy experimental hall or accelerated again by the superconducting linac (SCL2) in order to be used in the high energy experimental hall. In this paper, we will describe the results of the start-to-end simulations with the rare isotope beams generated by the ISOL system in the post accelerator of the RAON accelerator. In addition, the error analysis and correction at the superconducting linac SCL3 will be presented.
\end{abstract}



\maketitle

\section{INTRODUCTION}
\begin{figure}
\includegraphics[width=10.0cm]{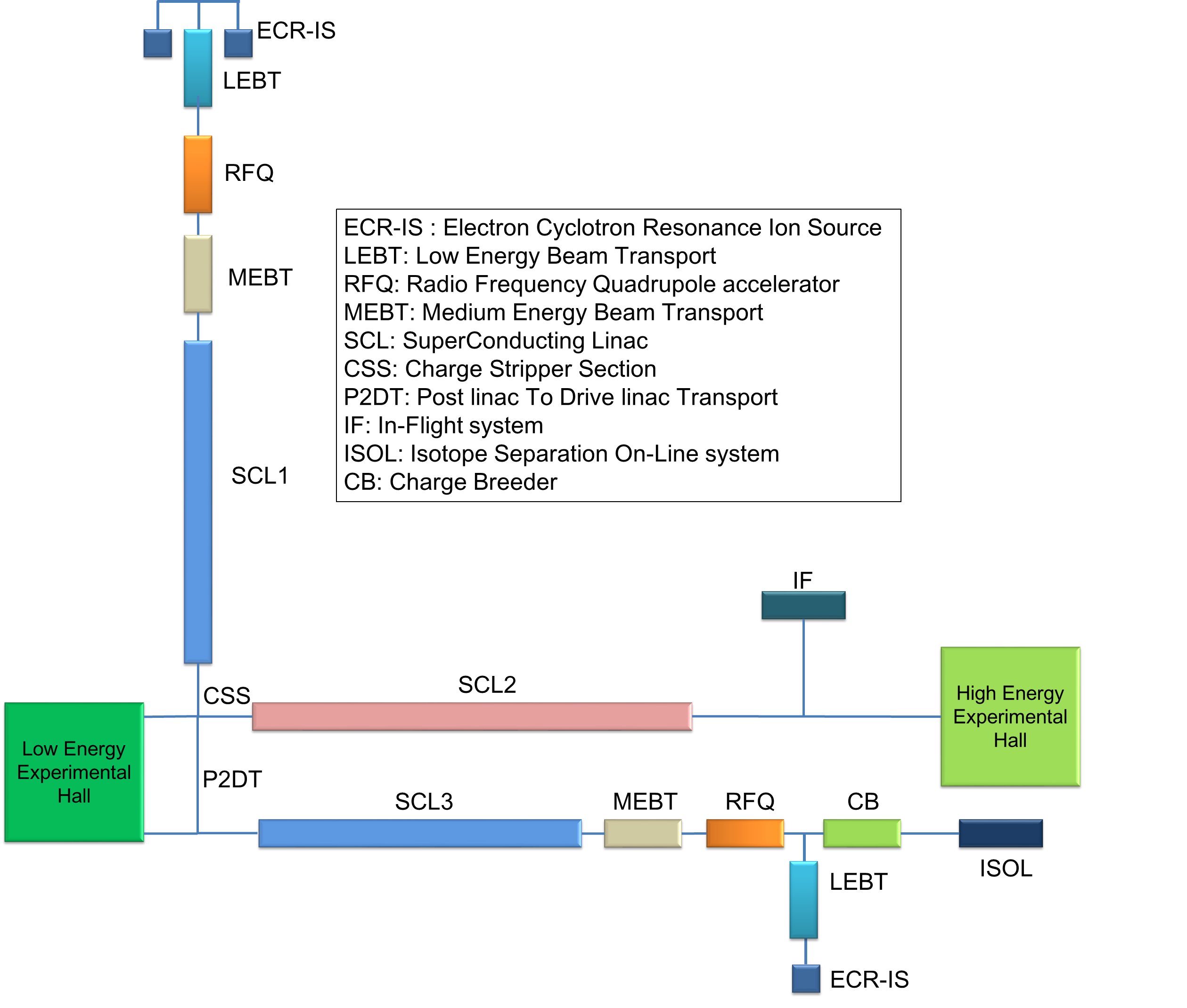}
\caption{Layout of the RAON accelerator.}
\label{RAON}
\end{figure}

The Rare Isotope Science Project (RISP) was established in December 2011 to construct RAON (Rare isotope Accelerator Of Newness) accelerator for various science programs. The project of the RAON accelerator~\cite{RAONbase, RAONJKPS} is in progress in order to generate and accelerate a variety of stable heavy ion beams and rare isotope beams to be used for a wide range of basic science researches and various applications. To produce various rare isotope beams, the RAON accelerator put to use the in-flight fragmentation (IF) system and the Isotope Separation On-Line (ISOL) system. At first, the IF system uses a diver linac, which consists of a 28 GHz superconducting electron cyclotron resonance ion source (ECR-IS), a main low energy beam transport (LEBT) section~\cite{LEBT}, a radio-frequency quadrupole (RFQ) accelerator, a medium energy beam transport (MEBT) section, a low energy superconducting linac (SCL1), a charge stripper section (CSS), and a high energy superconducting linac (SCL2). The beams accelerated by the SCL2 collide with the IF target, and then the rare isotope beams are created from the collision. Secondly, the ISOL system uses a 70 MeV cyclotron as the driver to deliver a 70 kW beam power up to the ISOL target. The rare isotope beams created by the ISOL system are accelerated again by a post accelerator, which consists of a post LEBT~\cite{postLEBT}, a post RFQ, a post MEBT, and a low energy superconducting linac SCL3. The beams accelerated by the SCL3 will be delivered up to the low energy experimental hall or to SCL2 after passing through the the post accelerator to the driver linac transport (P2DT) section~\cite{P2DT}. A schematic layout of RAON accelerator is shown in Fig.~\ref{RAON}.

\begin{figure}
\includegraphics[width=10.0cm]{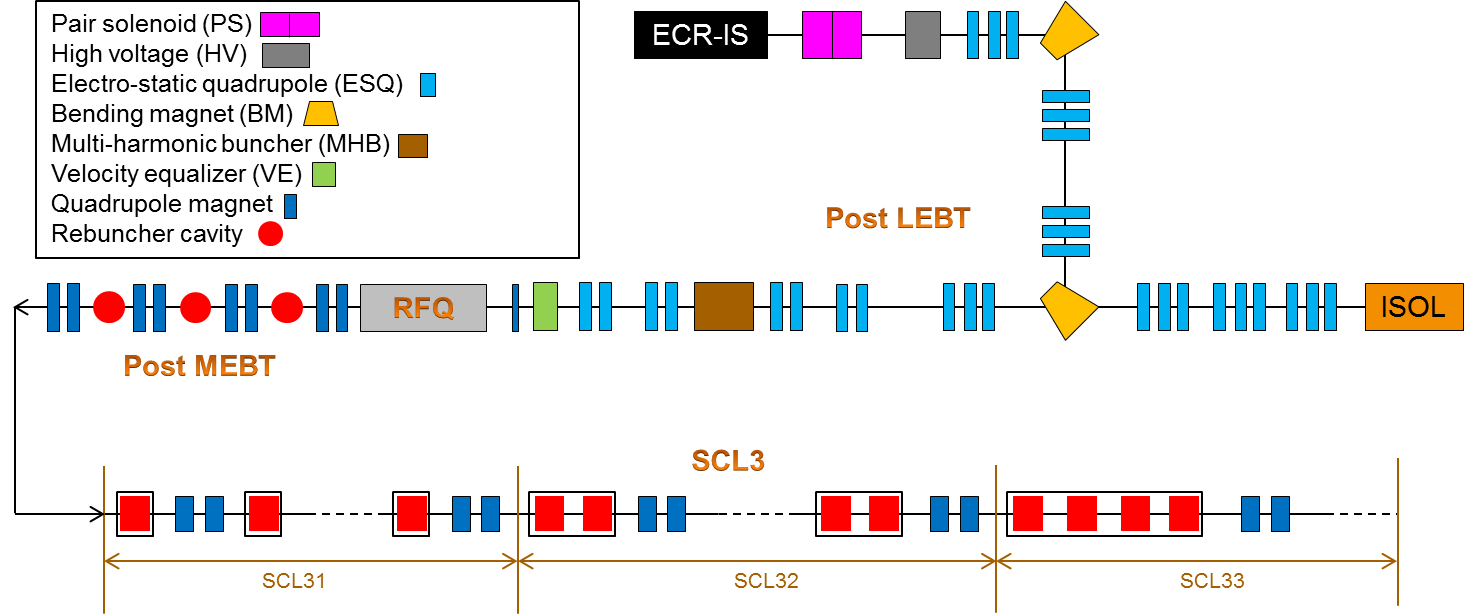}
\caption{Layout of the post accelerator in the RAON accelerator.}
\label{post_accelerator}
\end{figure}

As above mentioned, the post accelerator consists of four sections; the post LEBT, the post RFQ, the post MEBT, and the SCL3. First of all, the lattice of the post LEBT was partly modified because that of the ISOL system was recently changed. The new lattice design and beam dynamics simulations of the post LEBT were presented in ~\cite{postLEBT}. The main goal of the post LEBT is the stable transportation of the heavy ion beams and rare isotope beams up to the post RFQ. To accomplish this goal, the electro-static quadrupoles are used in the post LEBT. Secondly, the post RFQ of the post accelerator and the RFQ of the driver linac are the same, which is about 5 m long and the number of cell is 245. The maximum peak surface electric field is 17 MV/m and thus the post RFQ can accelerate 10 keV/u beam to 500 keV/u. Thirdly, the post MEBT includes 8 quadrupoles and 3 buncher cavities, which are used to satisfy the beam requirements of the SCL3. Finally, the SCL3 is divided into three sections, SCL31, SCL32 and SCL33, depending on the cavity type and the number of cavities in one cryomodule. The first section, the SCL31 is composed of 22 quarter-wave resonator (QWR) type cavities and each cavity is surrounded by one cryomodule. The second section, the SCL32 has 26 half-wave resonator (HWR) type cavities and one cryomodule surrounds two cavities. Third, the SCL33 has 76 HWR type cavities and one cryomodule surrounds 4 cavities. The schematic layout of the post accelerator is shown in Figure~\ref{post_accelerator}.

In this paper, we will present the results of the start-to-end simulations for the post accelerator with the reference beam $^{132}$Sn$^{33+}$. In addition, the study of the error analysis and correction at the superconducting linac SCL3 will be described by using the graphical user interface (GUI) based on the MATLAB program and the DYNAC~\cite{DYNAC} code in RAON accelerator~\cite{GUI}. 

\section{Start-to-end simulation}

\subsection{Beam information}

\begin{table}
\caption{Initial beam information.}
\begin{ruledtabular}
\begin{tabular}{llcc}
 Parameter & Value & Unit \\
\colrule
 Beam &  $^{132}$Sn$^{33+}$ & - \\
 Energy &  10.0 & keV/u \\
 Normalized transverse emittance &  0.1 & mm$\cdot$mrad \\
 $\beta_{0}$ &  0.2 & mm/mrad \\
 $\alpha_{0}$ &  0.1 & - \\
 Number of macro-particles &  100,000 & - \\
 Beam distribution &  4D water-bag & - \\
\end{tabular}
\end{ruledtabular}
\label{beam_info}
\end{table}

\begin{figure}
\includegraphics[width=10.0cm]{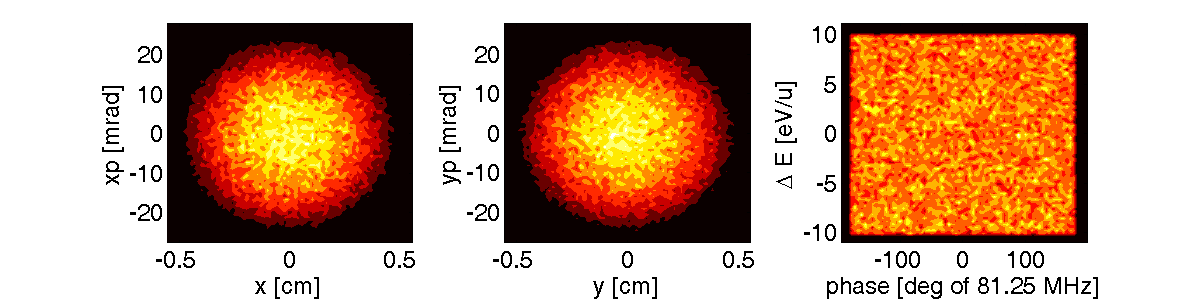}
\caption{Beam distributions in the transverse and longitudinal phase spaces.}
\label{beam_dist_ini}
\end{figure}

The reference beam generated by the ISOL system was recently determined as a tin beam, $^{132}$Sn$^{33+}$. At the end of the ISOL system, the beam energy is 10 keV/u and the normalized transverse emittance is 0.1 mm$\cdot$mrad. The basic beam information at the entrance of the post LEBT is listed in Table~\ref{beam_info}. Based on this beam information, we carry out the start-to-end simulation with 100,000 macro-particles using the particle tracking code, TRACK~\cite{TRACK}. Figure~\ref{beam_dist_ini} shows the beam distributions in the transverse and longitudinal phase spaces at the entrance of the post LEBT.

\subsection{Simulation results}

\begin{figure}
\includegraphics[width=10.0cm]{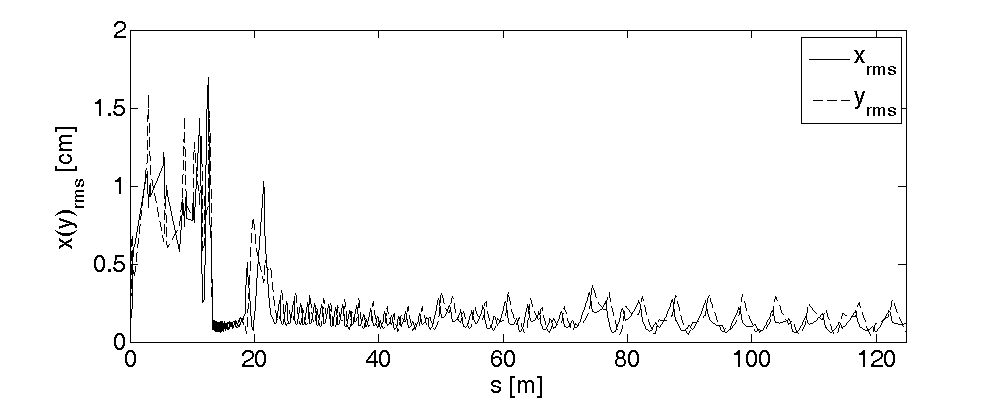}
\caption{Transverse rms beam size along the post accelerator.}
\label{rmsxy}
\end{figure}

Along the post accelerator, the beam pipe radius of each section is different, which is 6.0 cm at the post LEBT, 2.5 cm at the post MEBT, and 2.0 cm at the SCL3, respectively. Those beam pipe radii were determined by considering the beam energy and the optimization of the cavities. Therefore, the beam size should be less than the beam pipe radius to avoid the beam loss at each section. In view of such condition, the strength of each magnet was calculated and the start-to-end simulation was conducted. Figure~\ref{rmsxy} shows the transverse root-mean-square (rms) beam size along the post accelerator. Especially, the rms beam size is kept less than 0.5 cm at the SCL3 and the beam transmission after the post RFQ is about 98 \%, which is one of the advantages of the RAON accelerator.

\begin{figure}
\includegraphics[width=10.0cm]{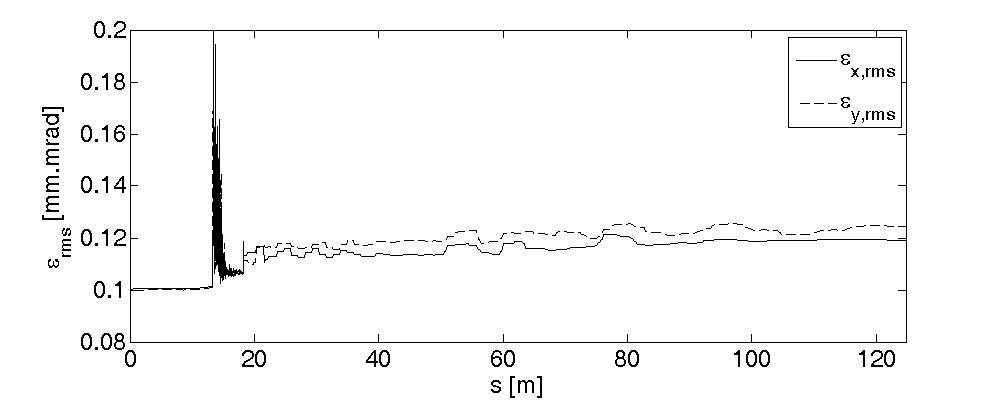}
\caption{Normalized transverse rms beam emittance along the post accelerator.}
\label{exyrn}
\end{figure}

Figure~\ref{exyrn} shows the normalized transverse rms beam emittance along the post accelerator. After passing through the RFQ, the energy spread becomes larger, which affects the transverse emittance growth. Also, the vertical emittance increases at the SCL3 because of the effect of the electro-magnetic field of the QWR cavity.

\begin{figure}
\includegraphics[width=10.0cm]{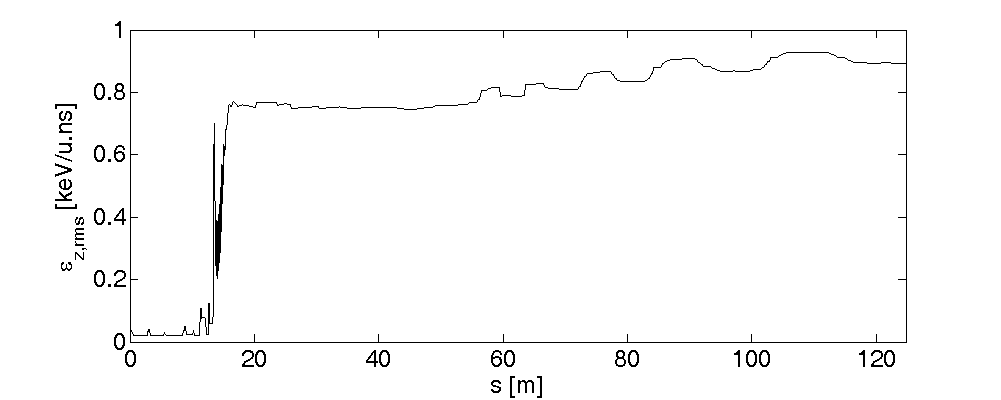}
\caption{Normalized longitudinal rms beam emittance along the post accelerator.}
\label{ezrn}
\end{figure}

\begin{table}
\caption{Normalized transverse and longitudinal rms emittance at some positions of the post accelerator.}
\begin{ruledtabular}
\begin{tabular}{llcc}
 Parameter & Horizontal [mm$\cdot$mrad] & Vertical [mm$\cdot$mrad] & Longitudinal [keV/u.ns] \\
\colrule
 End of RFQ   & 0.113  & 0.111   & 0.750 \\
 End of MEBT  & 0.113  & 0.116   & 0.758 \\
 End of SCL31 & 0.114  & 0.119   & 0.742 \\
 End of SCL32 & 0.117  & 0.120   & 0.856 \\
 End of SCL33 & 0.119  & 0.125   & 0.889 \\
\end{tabular}
\end{ruledtabular}
\label{emitt_table}
\end{table}

The normalized longitudinal rms beam emittance along the post accelerator is shown in Figure~\ref{ezrn}. After the beam is longitudinally bunched by the post RFQ, the energy spread increases because of the long tail of the beam and thus the longitudinal emittance increases. Table~\ref{emitt_table} shows the normalized transverse and longitudinal rms emittance at some position of the post accelerator. The growth of the horizontal, transverse and longitudinal emittances from the end of the RFQ to the end of the SCL33 is about 5.3 \%, 12.6 \% and 18.5 \%, respectively. The minimum requirement for the normalized transverse rms emittance at the low energy experimental halls is about 2 mm$\cdot$mrad which is much larger than the result of simulation.

\begin{figure}
\includegraphics[width=10.0cm]{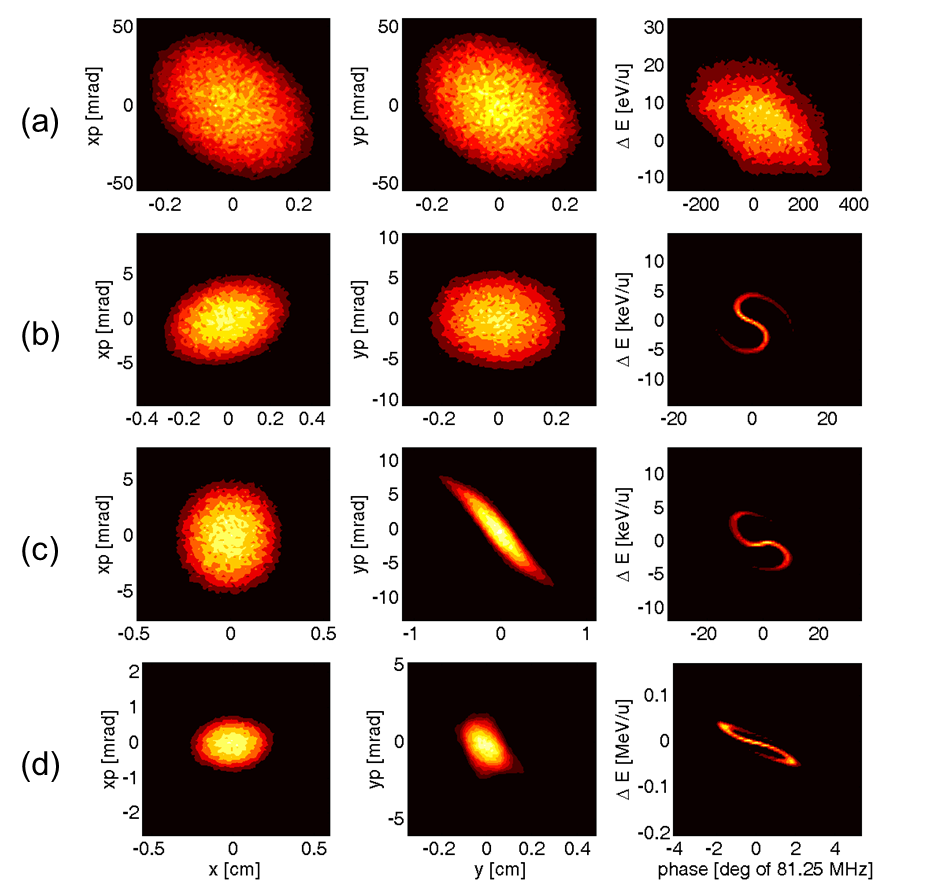}
\caption{Beam distribution in the phase space along the post accelerator: (a) post LEBT exit (b) post RFQ exit (c) post MEBT exit (d) SCL3 exit.}
\label{beam_dist}
\end{figure}

Figure~\ref{beam_dist} shows the transverse and longitudinal beam distributions at the end of each section. As mentioned above, the beam is longitudinally bunched by the post RFQ and it leads to the growth of the longitudinal beam emittance.

\section{Error simulations}
\subsection{Error tolerance}

\begin{table}
\caption{Tolerance of error sources at the SCL3. }
\begin{ruledtabular}
\begin{tabular}{llcc}
 - & - & Value & Unit \\
\colrule
 Initial position & x                       & +0.68/-0.65  & [cm]   \\
 Initial position & y                       & +0.27/-0.25  & [cm]   \\
 Initial angle    & xp                      & +8.0/-7.9    & [mrad] \\
 Initial angle    & yp                      & +4.0/-3.7    & [mrad] \\
 Cavity           & rms misalignment x,y    & 0.12         & [cm]   \\
 Cavity           & rms misalignment z      & 0.08         & [cm]   \\
 Cavity           & rms field amplitude     & 2.1          & [\%]   \\
 Cavity           & rms field phase         & 0.9          & [deg]  \\
 Quadrupole       & rms misalignment x,y    & 0.018        & [cm]   \\
 Quadrupole       & rms misalignment z      & 1.2          & [cm]   \\
 Quadrupole       & rms tilt                & 16.2         & [mrad] \\
\end{tabular}
\end{ruledtabular}
\label{error_tol}
\end{table}

In the SCL3, the beam orbit can be distorted by a variety of error sources like magnet misalignment, cavity field error and so on. After the beam loss caused by the orbit distortion is checked, the effect of each error source can be verified. In this paper, the tolerance of each error source is determined when the beam loss caused by each error source becomes smaller than 0.1 \%. The TRACK code is used for the calculation of the tolerance and the results are listed in Table~\ref{error_tol}. In the simulations, the errors of quadrupoles and cavities are given by Gaussian distribution with the rms value and truncated at the 3 times of the rms value. Among the error sources, the tolerance of the quadrupole transverse rms misalignment is about 0.018 cm and it is the most dominant one among all error sources.

\subsection{Orbit correction}

\begin{figure}
\includegraphics[width=10.0cm]{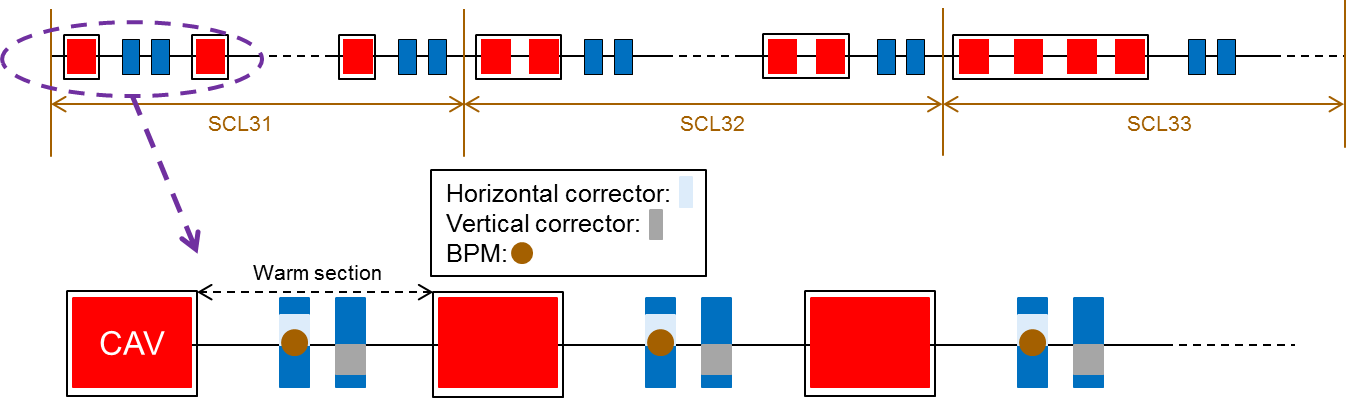}
\caption{Layout of the SCL3 in RAON accelerator.}
\label{SCL3}
\end{figure}

Figure~\ref{SCL3} shows the schematic layout of the SCL3. For the orbit correction, a horizontal corrector and a beam position monitor (BPM) are located at first quadrupole and a vertical corrector is located at second quadrupole at each warm section. With the correctors and BPMs, the orbit correction for the distorted orbit is carried out at the SCL3.

\begin{table}
\caption{Errors used in the orbit correction simulations.}
\begin{ruledtabular}
\begin{tabular}{lcccc}
 Parameter & Value & Unit \\
\colrule
 Initial position x,y & 10 & $\mu$m  \\
 Initial angle xp,yp  & 10 & $\mu$rad \\
 Cavity rms misalignment x,y     & 10 & $\mu$m \\
 Quadrupole rms misalignment x,y & 100 -- 500 & $\mu$m \\
\end{tabular}
\end{ruledtabular}
\label{error}
\end{table}

The errors used in the orbit correction simulations are listed in Table~\ref{error}. The beam orbit is distorted by these errors and then the orbit correction is carried out by using the correctors and BPMs with the singular value decomposition (SVD) method. The rms transverse misalignment of the quadrupole, which is the most dominant error source, changes from 100 $\mu$m to 500 $\mu$m.

\begin{figure}
\includegraphics[width=10.0cm]{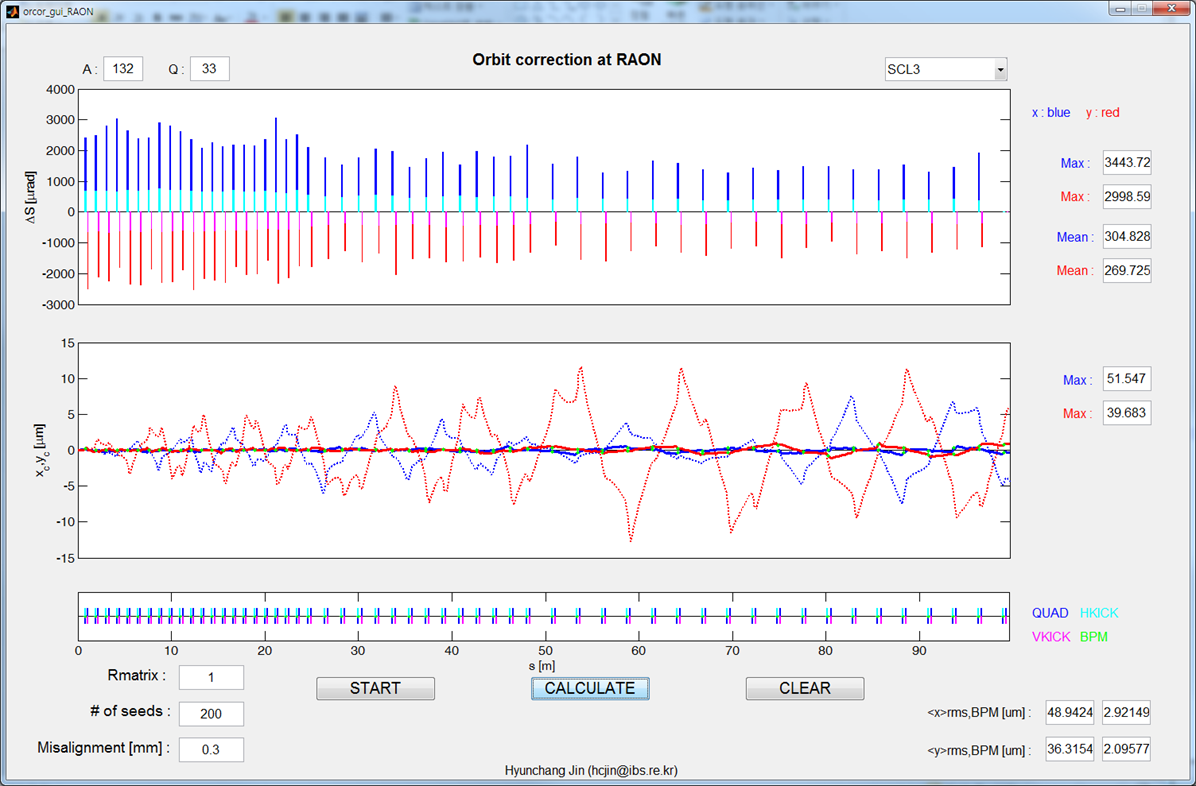}
\caption{GUI for the orbit correction in RAON accelerator with the quadrupole misalignment 300 $\mu$m and 200 random seeds.}
\label{gui_orbitcor}
\end{figure}

\begin{table}
\caption{Summary of the orbit correction for the quadrupole misalignment 300 $\mu$m and 200 random seeds.}
\begin{ruledtabular}
\begin{tabular}{lcc}
 Parameter & Value & Unit \\
\colrule
 $<x>_{rms,BPM}$ before correction & 48.9 & $\mu$m \\
 $<y>_{rms,BPM}$ before correction & 36.3 & $\mu$m \\
 $<x>_{rms,BPM}$ after correction  & 2.9  & $\mu$m \\
 $<y>_{rms,BPM}$ after correction  & 2.1  & $\mu$m \\
 Average horizontal corrector strength & 0.30  & mrad \\
 Average vertical corrector strength   & 0.27  & mrad \\
\end{tabular}
\end{ruledtabular}
\label{error_300}
\end{table}

For the orbit correction in RAON accelerator, the GUI based on the MATLAB program and the DYNAC code has been developed. Figure~\ref{gui_orbitcor} shows the GUI screen after the orbit correction with quadrupole misalignment 300 $\mu$m and 200 random seeds. As listed in Table~\ref{error_300}, the horizontal (vertical) rms beam size decreases from 48.9 (36.3) $\mu$m to 2.9 (2.1) $\mu$m after the orbit correction. The average corrector strength is less than the mechanical maximum value.

\begin{figure}
\includegraphics[width=10.0cm]{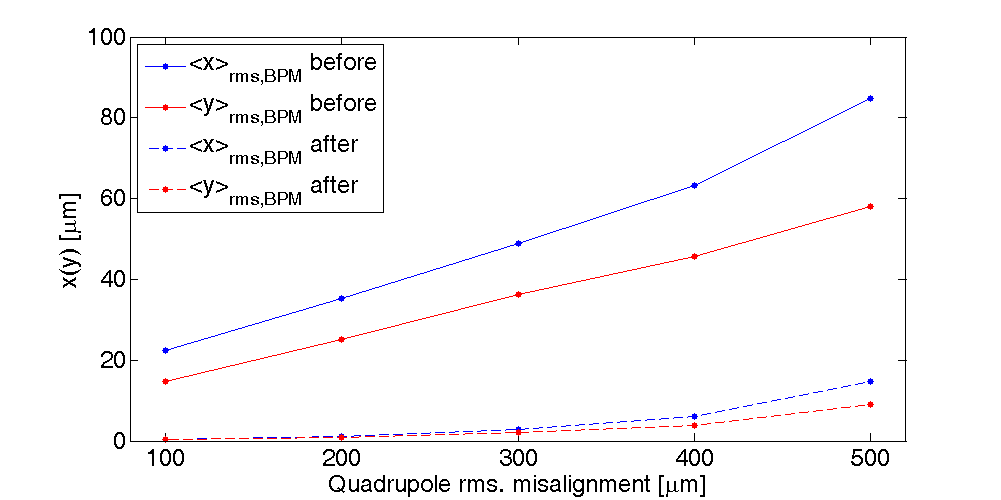}
\caption{Rms orbit size at BPMs before and after the orbit correction.}
\label{xy_orbitcor}
\end{figure}

\begin{figure}
\includegraphics[width=10.0cm]{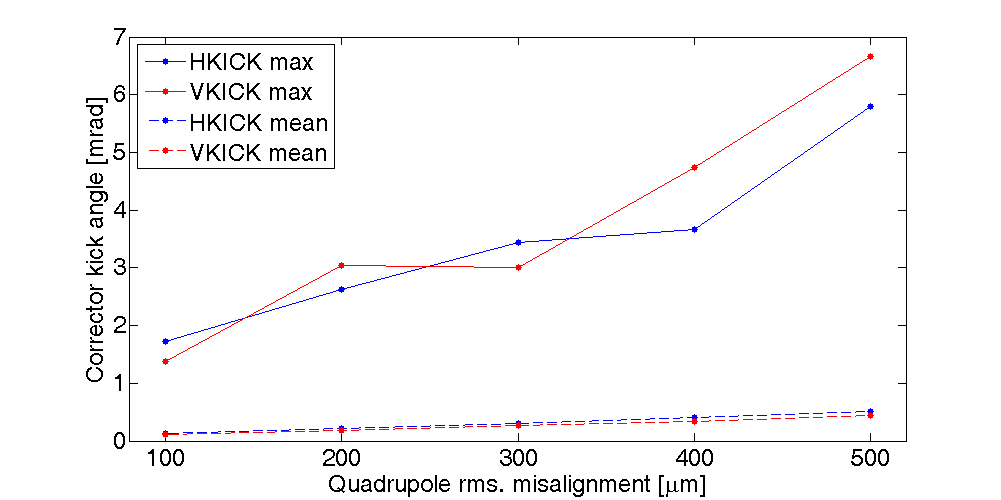}
\caption{Maximum and average corrector strength for the orbit correction.}
\label{corr_orbitcor}
\end{figure}

Figure~\ref{xy_orbitcor} shows the rms orbit size at the BPMs before and after the orbit correction for 100--500 $\mu$m rms quadrupole misalignment with 200 random seeds. For the quadrupole rms misalignment from 100 $\mu$m 500 $\mu$m, the rms orbit size decreases about -90 \%. The maximum and average corrector strength for the orbit correction is shown in Figure~\ref{corr_orbitcor}. For some random seeds, the corrector kick angle becomes larger, but the average value is less than 0.6 mrad which is much less than the mechanical maximum kick angle, about 2.7 mrad. The research for the proper number and position of the correctors and BPMs will be continued.

\section{CONCLUSIONS}
We presented the results of the start-to-end simulations with the rare isotope beam in the post accelerator of the RAON accelerator. For the post accelerator, the new reference beam, $^{132}$Sn$^{33+}$, from the ISOL system was tracked from the post LEBT to the superconducting linac SCL3. At the end of the SCL3, the beam energy reached at about 27.7 MeV/u and the rms beam size along the post accelerator was kept much less than the beam pipe radii. The error analysis and correction were also performed at the SCL3. The tolerance of each error source was calculated and the beam orbit distorted by the errors was corrected with the SVD method. Additionally, for various error sources, the orbit distortion was checked and corrected with the correctors and BPMs. As a result, the average kick angle of correctors for the orbit correction was much less than the mechanical maximum value. The research for the proper number and position of the correctors and BPMs will be carried out continuously.

\begin{acknowledgments}
This work was supported by the Rare Isotope Science Project of Institute for Basic Science funded by Ministry of Science, ICT and Future Planning and National Research Foundation of Korea (2013M7A1A1075764).
\end{acknowledgments}

\end{document}